\documentstyle[12pt]{article}
\begin{document}
\setcounter{page}{1}

%  *******************     New Commands     ******************

\newcommand{\be}{\begin{equation}}
\newcommand{\ee}{\end{equation}}
\newcommand{\beq}{\begin{eqnarray}}
\newcommand{\eeq}{\end{eqnarray}}

\newcommand{\eg}{{\it e.g.,\/}} %eg and ie should be written this
\newcommand{\ie}{{\it i.e.,\/}} %way.

\newcommand{\I}{{\rm \bf I}}
\newcommand{\J}{{\rm \bf J}}
\newcommand{\K}{{\rm \bf K}}

\newcommand{\ti}{\tilde}

\newcommand{\al}{\alpha}
\newcommand{\bt}{\beta}
\newcommand{\g}{\gamma}
\newcommand{\dl}{\delta}
\newcommand{\ep}{\epsilon}
\newcommand{\vep}{\varepsilon}
\newcommand{\z}{\zeta}
\newcommand{\e}{\eta}
\newcommand{\th}{\theta}
\newcommand{\vth}{\vartheta}
\newcommand{\io}{\iota}
\newcommand{\k}{\kappa}
\newcommand{\la}{\lambda}
\newcommand{\m}{\mu}
\newcommand{\n}{\nu}
\newcommand{\x}{\xi}
\newcommand{\p}{\pi}
\newcommand{\vp}{\varpi}
\newcommand{\r}{\rho}
\newcommand{\vr}{\varrho}
\newcommand{\s}{\sigma}
\newcommand{\vs}{\varsigma}

\newcommand{\up}{\upsilon}
\newcommand{\ph}{\phi}
\newcommand{\vph}{\varphi}

\newcommand{\ps}{\psi}
\newcommand{\om}{\omega}

\newcommand{\G}{\Gamma}
\newcommand{\D}{\Delta}
\newcommand{\Th}{\Theta}
\newcommand{\La}{\Lambda}
\newcommand{\X}{\Xi}

\newcommand{\Si}{\Sigma}
\newcommand{\U}{\Upsilon}
\newcommand{\Ph}{\Phi}
\newcommand{\Ps}{\Psi}
\newcommand{\Om}{\Omega}

\newcommand{\omp}{\omega^{+}}
\newcommand{\omn}{\omega^{-}}

%*****************************TitlePage*****************************
\begin{center}
\hfill ITP-SB-94-25 \\
\hfill hep-th/9406150\\
\vskip .75in

{\large SUPER W-SYMMETRIES, COVARIANTLY CONSTANT FORMS AND DUALITY
TRANSFORMATIONS}\\
\vskip .5in
Byungbae Kim\footnotemark
\footnotetext{e-mail address: bkim@insti.physics.sunysb.edu}  \\
Institute for Theoretical Physics, \\ State University of New York at Stony
Brook, \\ Stony Brook, NY 11794 USA \\

\vspace{4cm}
\end{center}

\begin{abstract}
On a supersymmetric sigma model the covariantly constant forms are related to
the conserved currents that are generators of a super W-algebra extending the
superconformal algebra. The existence of covariantly constant forms restricts
the holonomy group of the manifold. Via duality transformation we get new
covariantly constant forms, thus restricting the holonomy group of the new
manifold.
\end{abstract}

\vfill
\newpage
\section{Introduction}

One of the most intriguing properties of string theory is the existence of a
duality symmetry that relates strings propagating on different backgrounds to
each other (see, \eg \cite{Giveon} and references therein). Duality
transformations play an important role in the study of moduli spaces of string
vacua, and are presumably part of a large and fundamental symmetry of string
theory. The purpose of this note is to investigate how duality acts on $N=1$
sigma models with torsion possessing an extended algebra of conserved currents,
\ie a super $W$-algebra. As was shown in \cite{Howe}, conserved chiral
currents can in a natural way be associated to covariantly closed forms on the
target manifold. Since covariantly closed forms imply certain constraints on
the holonomy, the manifolds will in general have reduced holonomy. A simple
example of such a manifold is a Calabi Yau manifold, which has reduced holonomy
SU(3) and whose covariantly constant complex structure enhances the
superconformal algebra to an N=2 algebra. In this letter we show explicitly how
conserved currents associated to covariantly constant differential forms are
under duality mapped to another set of conserved currents on the dual manifold.
It will turn out that these latter currents are also associated to covariantly
constant forms on the dual manifold, for which we present explicit expressions.
{}From this we conclude that duality does, in general, leave the holonomy of
the
manifold invariant\footnote{Actually, as we shall see below, duality
transformations make use of an isometry of the manifold, and the holonomy is
preserved only if the isometry is compatible with the (covariantly constant)
forms.}.

\section{Super {\em W}-Algebra and Covariantly Constant Forms}

Given a supersymmetric nonlinear $\sigma$-model lagrangian
\be
S= -\frac{1}{2}\int D^{2}(g_{\mu \nu}D_{+}X^{\mu} D_{-}X^{\nu}),
\ee
for $J_{\om}=\om_{\s_{1}\cdots \s_{p}}D_{+}X^{\s_{1}}\cdots D_{+}X^{\s_{p}}$
\\ \\

\beq
D_{-}J_{\om} &=& \partial _{\m}\om_{\s_{1}\cdots \s_{p}}D_{-}X^{\m}
D_{+}X^{\s_{1}}\cdots D_{+}X^{\s_{p}} \nonumber \\ & &
+p\om_{\s_{1}\s_{2}\cdots \s_{p}}D_{-}D_{+}X^{\s_{1}} D_{+}X^{\s_{2}}\cdots
D_{+}X^{\s_{p}} \nonumber \\ &=& \partial _{\m}\om_{\s_{1}\cdots
\s_{p}}D_{-}X^{\m} D_{+}X^{\s_{1}}\cdots D_{+}X^{\s_{p}} \nonumber \\ & &
-p\om_{\s_{1}\s_{2}\cdots \s_{p}} \G^{\s_{1}}_{\r\s}D_{-}X^{\r}D_{+}X^{\s}
D_{+}X^{\s_{2}}\cdots D_{+}X^{\s_{p}} \nonumber \\ &=&
(\nabla_{\m}\om_{\s_{1}\cdots \s_{p}}) D_{-}X^{\m}D_{+}X^{\s_{1}}\cdots
D_{+}X^{\s_{p}}
\eeq
(where we used the equation of motion
$D_{-}D_{+}X^{\n}+\G^{\n}_{\r\s}D_{-}X^{\r}D_{+}X^{\s}=0$). \\

So $D_{-}J_{\om}=0$ if the form $\om$ is covariantly constant, \ie
$\nabla_{\m}\om =0$.  In this case $J_{\om}$ is a conserved current that is a
generator of a super $W$-algebra via Poisson Brackets, extending the
superconformal algebra \cite{Howe}. \\

The existence of such covariantly constant forms is related to the holonomy
group of the target manifold.  The following table shows the existence of
covariantly constant forms and the holonomy group of the manifold (for the
nonsymmetric spaces)\cite{Berger}.  \\

\begin{center}
{\bf Holonomy Groups and Covariantly Constant Forms on Nonsymmetric
spaces}\footnotemark
\footnotetext{The product of the algebra involves not only wedge products of
     forms but also contractions of two or more forms using the metric.} \\
\end{center}
\medskip
\begin{tabular}{||c|c|c|c||} \hline\hline
dim & holonomy group & covariantly constant forms & \\ & & (generators of the
algebra) & \\ \hline\hline $n$ & $SO(n)$ & $\om_{n}=$ volume form & \\ \hline
$2n$ & $U(n)$ & $(\om_{2} )^{k}$ $k=1,\cdots n$; & K\"{a}hler manifold \\ & &
$\om_{2}$: K\"{a}hler form & \\ \hline $2n$ & $SU(n)$ & $(\om_{2} )^{k}$
$k=1,\cdots n$; $\om_{n}=(vol)^ {\frac{1}{2}}$ & Special K\"{a}hler \\ & &
$\om_{n}$: complex volume form, $\bar{\om}_{n}$ & (Ricci flat K\"{a}hler) \\
\hline $4n$ & $Sp(n)$ & $\om =a\om_{1} +b\om_{2} +c\om_{3}$, & Hyperk\"{a}hler
\\ & & $a^{2}+b^{2}+c^{2}=1$ & \\ \hline $4n$ & $Sp(n)Sp(1)$ &
$\Om_{4}=\om_{1}^{2}+\om_{2}^{2}+\om_{3}^{2}$ & Quaternionic K\"{a}hler \\
\hline 7 & $G_{2}$ & $\om_{3},\om_{4}$ & \\ \hline 8 & $Spin(7)$ & $\om_{4}$ &
\\ \hline\hline
\end{tabular}
\bigskip

For each holonomy group, there is an associated covariantly constant
totally antisymmetric tensor, and this implies the existence of an associated
symmetry of the corresponding supersymmetric sigma model \cite{Howe}. These
currents generate a super W-algebra via Poisson Brackets and these algebras are
extensions of the superconformal algebra.

\vspace{1in}

\section{Covariantly Constant Forms and Duality Transformation}
Now given a supersymmetric nonlinear $\sigma$-model lagrangian (we allow
torsion for more generality) with covariantly constant $p$-forms $\om^\pm$ (see
below), we have the action
\be
S= -\frac{1}{2}\int D^{2}((g_{\mu \nu}+b_{\mu \nu})D_{+}X^{\mu} D_{-}X^{\nu})\
{}.
\label{eqor}
\ee
Suppose further there is an isometry generated by a vector field $Y$ with
${\cal L}_{Y}(db)=0$ and ${\cal L}_{Y}\om^{\pm}=0$; then one can choose (local)
coordinates such that $Y=\partial /\partial X^{0}$ and the metric $g$, the
torsion potential $b$, and the $p$-forms $\om^\pm$ are
independent of $X^{0}$.\\

Under the transformations
\be
(\dl X^{\m})g_{\m\n}=\ep \omp_{\n \s_{1}\cdots \s_{p-1}}D_{+}X^{\s_{1}}\cdots
D_{+}X^{\s_{p-1}}\ ,
\ee
\ie
\be
\dl X^{\m}=\ep g^{\m\n}\omp_{\n \s_{1}\cdots \s_{p-1}}D_{+}X^{\s_{1}}\cdots
                     D_{+}X^{\s_{p-1}}
\ee
(when $p=2$,
\beq
\dl X^{\m} &=& \ep g^{\m\n} \omp_{\n \s}D_{+}X^{\s}  \nonumber  \\
           &=& \ep J^{+\m}_{\s}D_{+}X^{\s}
\eeq
where $\omp _{\n\s}=g_{\n\r}J^{+\r}_{\s}$, the 2 form associated with the
complex structure $J^{+}$\cite{Bkim}), \\ the variation of the Lagrangian $L$
is
\beq
\dl L &=& 2g_{\m\n}\dl X^{\m}[D_{-}D_{+}X^{\n}+\G^{+\n}_{\r\s}D_{-}X^{\r}
                               D_{+}X^{\s}] \nonumber \\ &=& 2\ep \omp
_{\r\s_{1}\cdots \s_{p-1}}D_{+}X^{\s_{1}}\cdots
D_{+}X^{\s_{p-1}}[D_{-}D_{+}X^{\r}+\G^{+\r}_{\m\n}D_{-}X^{\m} D_{+}X^{\n}]
\nonumber \\ &=& 2\ep \frac{1}{p}[D_{-}(\omp_{\r\s_{1}\cdots \s_{p-1}}
D_{+}X^{\r}D_{+}X^{\s_{1}}\cdots D_{+}X^{\s_{p-1}}) \nonumber \\ & &
-(\nabla^{+}_{\m}\omp_{\r\s_{1}\cdots \s_{p-1}})
D_{-}X^{\m}D_{+}X^{\r}D_{+}X^{\s_{1}}\cdots D_{+}X^{\s_{p-1}}],
\eeq
where
\be
\G^{+\n}_{\r\s}=\G^{\n}_{\r\s}+T^{\n}_{\r\s}
\ee
and $T_{\m\n\r}=\frac{1}{2}(b_{\m\n ,\r}+b_{\n\r ,\m}+b_{\r\m ,\n})$
\cite{Gates,Rocek}. Similarly,
\beq
\dl L &=&   2\e \frac{1}{p}[D_{+}(\omn_{\r\s_{1}\cdots \s_{p-1}}
          D_{-}X^{\r}D_{-}X^{\s_{1}}\cdots D_{-}X^{\s_{p-1}}) \nonumber \\ & &
-(\nabla^{-}_{\m}\omn_{\r\s_{1}\cdots \s_{p-1}})
D_{+}X^{\m}D_{-}X^{\r}D_{-}X^{\s_{1}}\cdots D_{-}X^{\s_{p-1}}]
\eeq
(where $\nabla^-=\partial +\G^-$ and
$\G^{-\n}_{\r\s}=\G^{\n}_{\r\s}-T^{\n}_{\r\s}$)
under the transformation
\be
(\dl X^{\m})g_{\m\n}=\e \omn_{\n \s_{1}\cdots \s_{p-1}}D_{-}X^{\s_{1}}\cdots
D_{-}X^{\s_{p-1}}\ ,
\ee
\ie
\be
\dl X^{\m}=\e g^{\m\n}\omn_{\n \s_{1}\cdots \s_{p-1}}D_{-}X^{\s_{1}}\cdots
                     D_{-}X^{\s_{p-1}}.
\ee
\\

So the action is invariant if we choose the form $\om $ covariantly constant,
\ie $\nabla^{\pm}_{\m}\om^{\pm}_{\r\s_{1}\s_{p-1}}=0$.  \\

In this case
\be
J_{\om^{+}}=\om^{+}_{\r\s_{1}\cdots \s_{p-1}} D_{+}X^{\r}D_{+}X^{\s_{1}}\cdots
D_{+}X^{\s_{p-1}}
\ee
and
\be
J_{\om^{-}}=\om^{-}_{\r\s_{1}\cdots \s_{p-1}} D_{-}X^{\r}D_{-}X^{\s_{1}}\cdots
D_{-}X^{\s_{p-1}}
\ee
are conserved (Noether) currents that are generators of a super $W$-algebra
extending the superconformal algebra \cite{Howe}.  \\

\bigskip

Now one can get a new target manifold with new metric $\tilde{g}$ and torsion
potential $\tilde{b}$ by dualizing as in \cite{Fradkin,Hitchin,Bkim}.  Gauging
the symmetry with gauge fields $V_{\pm}$ (replace $D_{\pm}X^{0}$ with
$D_{\pm}X^{0}+V_{\pm}$ and choosing $X^{0}=0$ gauge), we get the first order
lagrangian
\begin{eqnarray}
S & = & -1/2\int D^{2}( e_{00}V_{+}V_{-}+e_{i0}D_{+}X^{i}V_{-}
+e_{0i}V_{+}D_{-}X^{i} +e_{ij}D_{+}X^{i}D_{-}X^{j} \nonumber \\ & & +\phi
(D_{+}V_{-}+D_{-}V_{+}) ) \label{eq1}
\end{eqnarray}
Here $\{\mu ,\nu\} =0,1,..,2n-1,\;\;\{i,j\}=1,..,2n-1,\;\;$
$e_{\mu\nu}=g_{\mu\nu}+b_{\mu\nu}$, and $\phi $ is the lagrange multiplier
whose variation gives $V_{\pm}=D_{\pm}X^{0}$ and gives back the original action
(\ref{eqor}).

The first order action (\ref{eq1}) is invariant under
\beq
\dl_{\ep}\ph = \ep e_{\m 0}g^{\m\n}[& &\omp_{\n 0\s_{2}\cdots \s_{p-1}}
                 V_{+}D_{+}X^{\s_{2}}\cdots D_{+}X^{\s_{p-1}} \nonumber \\ & &
+\cdots \;\;\;\;\;\;\; \cdots \nonumber \\ & & \omp_{\n\s_{1}\cdots
\s_{p-2}0}D_{+}X^{\s_{1}} \cdots D_{+}X^{\s_{p-2}}V_{+} \nonumber \\ & &
+\omp_{\n k_{1}\cdots k_{p-1}}D_{+}X^{k_{1}}\cdots D_{+}X^{k_{p-1}}],
\eeq
\beq
\dl_{\ep}X^{j} = \ep g^{j\n}[& &\omp_{\n 0\s_{2}\cdots \s_{p-1}}
                 V_{+}D_{+}X^{\s_{2}}\cdots D_{+}X^{\s_{p-1}} \nonumber \\ & &
+\cdots \;\;\;\;\;\;\; \cdots \nonumber \\ & & \omp_{\n\s_{1}\cdots
\s_{p-2}0}D_{+}X^{\s_{1}} \cdots D_{+}X^{\s_{p-2}}V_{+} \nonumber \\ & &
+\omp_{\n k_{1}\cdots k_{p-1}}D_{+}X^{k_{1}}\cdots D_{+}X^{k_{p-1}}],
\eeq
and
\beq
\dl_{\e}\ph = -\e e_{0\m}g^{\m\n}[& &\omn_{\n 0\s_{2}\cdots \s_{p-1}}
                 V_{-}D_{-}X^{\s_{2}}\cdots D_{-}X^{\s_{p-1}} \nonumber \\ & &
+\cdots \;\;\;\;\;\;\; \cdots \nonumber \\ & & \omn_{\n\s_{1}\cdots
\s_{p-2}0}D_{-}X^{\s_{1}} \cdots D_{-}X^{\s_{p-2}}V_{-} \nonumber \\ & &
+\omn_{\n k_{1}\cdots k_{p-1}}D_{-}X^{k_{1}}\cdots D_{-}X^{k_{p-1}}],
\eeq
\beq
\dl_{\e}X^{j} = \e g^{j\n}[& &\omn_{\n 0\s_{2}\cdots \s_{p-1}}
                 V_{-}D_{-}X^{\s_{2}}\cdots D_{-}X^{\s_{p-1}} \nonumber \\ & &
+\cdots \;\;\;\;\;\;\; \cdots \nonumber \\ & & \omn_{\n\s_{1}\cdots
\s_{p-2}0}D_{-}X^{\s_{1}} \cdots D_{-}X^{\s_{p-2}}V_{-} \nonumber \\ & &
+\omn_{\n k_{1}\cdots k_{p-1}}D_{-}X^{k_{1}}\cdots D_{-}X^{k_{p-1}}].
\eeq
\\

To find the dual model, we eliminate $V_{\pm}$ by the equations of motion and
get
\be
V_{+}= e_{00}^{-1}(D_{+}\phi -e_{i0}D_{+}X^{i}) \;\;\;\;\;\;
V_{-}=-e_{00}^{-1}(D_{-}\phi +e_{0i}D_{-}X^{i}) \label{eq2}
\ee
so that
\beq
\dl_{\ep}\ph = \ep e_{\m 0}g^{\m\n}[& &(p-1)e^{-1}_{00}\omp_{\n 0 k_{2}\cdots
                    k_{p-1}}D_{+}\ph D_{+}X^{k_{2}}\cdots D_{+}X^{k_{p-1}}
\nonumber \\ & & +(\omp_{\n k_{1}\cdots k_{p-1}}-(p-1)e^{-1}_{00}e_{k_{1}0}
\omp_{\n 0 k_{2}\cdots k_{p-1}})\times \nonumber \\ & &
D_{+}X^{k_{1}}D_{+}X^{k_{2}}\cdots D_{+}X^{k_{p-1}}],
\eeq
\beq
\dl_{\ep}X^{j} = \ep g^{j\n}[& &(p-1)e^{-1}_{00}\omp_{\n 0 k_{2}\cdots
                    k_{p-1}}D_{+}\ph D_{+}X^{k_{2}}\cdots D_{+}X^{k_{p-1}}
\nonumber \\ & & +(\omp_{\n k_{1}\cdots k_{p-1}}-(p-1)e^{-1}_{00}e_{k_{1}0}
\omp_{\n 0 k_{2}\cdots k_{p-1}})\times \nonumber \\ & &
D_{+}X^{k_{1}}D_{+}X^{k_{2}}\cdots D_{+}X^{k_{p-1}}],
\eeq
and
\beq
\dl_{\e}\ph = -\e e_{0\m}g^{\m\n}[& &-(p-1)e^{-1}_{00}\omn_{\n 0 k_{2}\cdots
                    k_{p-1}}D_{-}\ph D_{-}X^{k_{2}}\cdots D_{-}X^{k_{p-1}}
\nonumber \\ & & +(\omn_{\n k_{1}\cdots k_{p-1}}-(p-1)e^{-1}_{00}e_{0k_{1}}
\omn_{\n 0 k_{2}\cdots k_{p-1}})\times \nonumber \\ & &
D_{-}X^{k_{1}}D_{-}X^{k_{2}}\cdots D_{-}X^{k_{p-1}}],
\eeq
\beq
\dl_{\e}X^{j} = \e g^{j\n}[& &-(p-1)e^{-1}_{00}\omn_{\n 0 k_{2}\cdots
                    k_{p-1}}D_{-}\ph D_{-}X^{k_{2}}\cdots D_{-}X^{k_{p-1}}
\nonumber \\ & & +(\omn_{\n k_{1}\cdots k_{p-1}}-(p-1)e^{-1}_{00}e_{0k_{1}}
\omn_{\n 0 k_{2}\cdots k_{p-1}})\times \nonumber \\ & &
D_{-}X^{k_{1}}D_{-}X^{k_{2}}\cdots D_{-}X^{k_{p-1}}].
\eeq
and we can read off the dual covariantly constant forms $\tilde{\om}^{+}$ and
$\tilde{\om}^{-}$:
\be
\tilde{\om}^{+}_{0 k_{1}\cdots k_{p-1}}=g^{-1}_{00}\omp_{0 k_{1}\cdots
             k_{p-1}}
\ee
\be
\tilde{\om}^{+}_{jk_{1}\cdots k_{p-1}}=\omp_{jk_{1}\cdots k_{p-1}}
            -g^{-1}_{00}W^{+}_{jk_{1}\cdots k_{p-1}},
\ee
where
\beq
W^{+}_{jk_{1}\cdots k_{p-1}} &=& e_{j0}\omp_{0k_{1}\cdots k_{p-1}}
-e_{k_{1}0}\omp_{0jk_{2}\cdots k_{p-1}} \nonumber \\ & &
+e_{k_{2}0}\omp_{0k_{1}jk_{3}\cdots k_{p-1}} \nonumber \\ & & \cdots \nonumber
\\ & & +(-1)^{p-1}e_{k_{p-1}0}\omp_{0k_{1}\cdots k_{p-2}j},
\eeq
and
\be
\tilde{\om}^{-}_{0 k_{1}\cdots k_{p-1}}=-g^{-1}_{00}\omn_{0 k_{1}\cdots
             k_{p-1}}
\ee
\be
\tilde{\om}^{-}_{jk_{1}\cdots k_{p-1}}=\omn_{jk_{1}\cdots k_{p-1}}
            -g^{-1}_{00}W^{-}_{jk_{1}\cdots k_{p-1}},
\ee
where
\beq
W^{-}_{jk_{1}\cdots k_{p-1}} &=& e_{0j}\omn_{0k_{1}\cdots k_{p-1}}
-e_{0k_{1}}\omn_{0jk_{2}\cdots k_{p-1}} \nonumber \\ & &
+e_{0k_{2}}\omn_{0k_{1}jk_{3}\cdots k_{p-1}} \nonumber \\ & & \cdots \nonumber
\\ & & +(-1)^{p-1}e_{0k_{p-1}}\omn_{0k_{1}\cdots k_{p-2}j}.
\eeq
\\

Now the new currents in the dual space are (with $X^{0}=\ph$)
\be
\tilde{J}_{\tilde{\om}^{+}}=\tilde{\om}^{+}_{\r\s_{1}\cdots \s_{p-1}}
             D_{+}X^{\r}D_{+}X^{\s_{1}}\cdots D_{+}X^{\s_{p-1}}
\ee
and
\be
\tilde{J}_{\tilde{\om}^{-}}=\tilde{\om}^{-}_{\r\s_{1}\cdots \s_{p-1}}
             D_{-}X^{\r}D_{-}X^{\s_{1}}\cdots D_{-}X^{\s_{p-1}}
\ee
which generate a super $W$-algebra in the new setting.  \\

Also substituting (\ref{eq2}) into (\ref{eq1}), we get the new lagrangian
\cite{Fradkin,Buscher}
\beq
\tilde S=-\frac12\int D^2\left( e^{-1}_{00}D_{+}\phi D_{-}\phi
+e^{-1}_{00}e_{0j}D_{+}\phi D_{-}X^{j}
-e^{-1}_{00}e_{i0}D_{+}X^{i}D_{-}\phi\right.\nonumber\cr
\left. +(e_{ij}-e^{-1}_{00}e_{i0}e_{0j})
D_{+}X^{i}D_{-}X^{j}\right)
\eeq
and read off the dual metric \~{g} and torsion potential \~{b}, \ie
\be
\tilde{g}_{\mu \nu}=
\left( \begin{array}{cc}
 g^{-1}_{00} & g^{-1}_{00}b_{0j} \\ g^{-1}_{00}b_{0i} &
g_{ij}-g^{-1}_{00}g_{i0}g_{j0}+g^{-1}_{00}b_{i0}b_{j0} \end{array} \right),
\ee
and
\be
\tilde{b}_{\mu \nu}=
\left( \begin{array}{cc}
  0 & g^{-1}_{00}g_{0j} \\ -g^{-1}_{00}g_{0i} &
b_{ij}+g^{-1}_{00}(g_{i0}b_{j0}-g_{j0}b_{i0}) \end{array} \right)
\ee
with respect to the basis $\{d\phi,\;dX^{1}\;..\;dX^{2n-1}\}$.  \\

\section{Conclusions}
The existence of covariantly constant forms restricts the holonomy group of the
manifold. When the connection is the Levi-civita connection, the classification
was done by Berger \cite{Berger}. In the case of sigma model with a riemannian
target manifold, via duality transformations we get another manifold with new
metric and torsion and the existence of the covariantly constant forms will
restrict the holonomy of this new manifold.  The classification of holonomy
groups in the case of the connections with torsion, generalizing Berger's work,
is an interesting area to be studied and the duality transformation will be
very useful for that purpose.  \\

\bigskip

\noindent
{\bf Acknowledgments}: \\

\medskip

I thank Martin Ro\v cek for suggesting this problem
and for many discussions.  I also thank Jan de Boer for helpful comments.
Finally, I thank NSF Grant No.\ PHY 93-09888 for partial support.

\end{document}